\documentclass[12pt,preprint]{aastex}

\newcommand{\ms}{M$_{\odot}$}

\newcommand{\hh}{H$_2$}
\newcommand{\kms}{$\,\rm km\,s^{-1}$}

\newcommand{\kmspc}{$\rm km\,s^{-1}\,pc^{-1}$}

\newcommand{\be}{\begin{equation}}
\newcommand{\ee}{\end{equation}}
\newcommand{\dvdz}{$\rm dv/dz$}

\newcommand{\as}{$^{\prime\prime}$}
\newcommand{\ccm}{\,\mathrm{cm^{-3}}}  

\newcommand{\too}{\!\rightarrow\!} 
\newcommand{\jone}{\small{$\rm J\eqq1\too0$}\normalsize}
\newcommand{\jtwo}{\small{$\rm J\eqq2\too1$}\normalsize}
\newcommand{\jthree}{\small{$\rm J\eqq3\too2$}\normalsize}
\newcommand{\jfour}{\small{$\rm J\eqq4\too3$}\normalsize}
\newcommand{\jfive}{\small{$\rm J\eqq5\too4$}\normalsize}
\newcommand{\jsix}{\small{$\rm J\eqq6\too5$}\normalsize}
\newcommand{\jseven}{\small{$\rm J\eqq7\too6$}\normalsize}
\newcommand{\jsixteen}{\small{$\rm J\eqq16\too15$}\normalsize}
\newcommand{\ncodv}{$\rm N_{CO} / dv$}
\newcommand{\xco}{$\rm X_{CO}$}

\newcommand{\by}{$\times$}
\newcommand{\e}{\times10^}

\newcommand{\eqq}{\!=\!} 

\slugcomment{Accepted for publication in The Astrophysical Journal}

\shorttitle{Cosmic-ray heated molecular gas in NGC~253}
\shortauthors{Bradford et al.}

\begin{document}

\title{CO ($\rm J=7\rightarrow 6$) Observations of NGC 253: Cosmic Ray Heated Warm Molecular 
Gas}

\author{
 C.M. Bradford\altaffilmark{1,2},
 T. Nikola\altaffilmark{1},
 G.J. Stacey\altaffilmark{1},
 A.D. Bolatto\altaffilmark{3},
 J.M. Jackson\altaffilmark{3,4},
 M.L. Savage\altaffilmark{5}, 
 J.A. Davidson\altaffilmark{5}, and
 S.J. Higdon\altaffilmark{6,1}
}

\altaffiltext{1}{Department of Astronomy, Cornell Univ., Ithaca, NY 14853}
 \altaffiltext{2}{Division of Physics, Math and Astronomy, Cal. Inst. of
 Tech., Pasadena, CA 91125} 
\altaffiltext{3}{Inst. for Astrophysical Research,
 Boston Univ., Boston, MA 02215} 
\altaffiltext{4}{Radio Astronomy Lab., U.C. Berkeley, Berkeley, CA 94720} 
\altaffiltext{5}{USRA/NASA Ames Research Ctr., Moffett
 Field, CA 94035} 
\altaffiltext{6}{Physics Department, Queen Mary \& Westfield College,
 Univ. of London, London E1 4NS, UK}

\begin{abstract}

We report observations of the CO \jseven\ transition toward the starburst nucleus of
NGC~253.  This is the highest-excitation CO measurement in this source to date, and allows
an estimate of the molecular gas excitation conditions.  Comparison of the CO line
intensities with a large velocity gradient, escape probability model indicates that the
bulk of the 2--5$\rm\times10^{7}~M_{\odot}$ of molecular gas in the central 180 pc is
highly excited.  A model with T~$\sim$~120~K, $\rm n_{H_2}\sim~4.5\e{4}~cm^{-3}$ is
consistent with the observed CO intensities as well as the rotational \hh\ lines observed
with ISO.

The inferred mass of warm, dense molecular gas is 10--30 times the atomic gas mass as traced
through its [\ion{C}{2}] and [\ion{O}{1}] line emission.  This large mass ratio is
inconsistent with photodissociation region models where the gas is heated by far-UV
starlight.  It is also not likely that the gas is heated by shocks in outflows or
cloud-cloud collisions.  We conclude that the best mechanism for heating the gas is cosmic
rays, which provide a natural means of uniformly heating the full volume of molecular
clouds.  With the tremendous supernova rate in the nucleus of NGC~253, the CR heating rate
is at least $\sim$ 800 times greater than in the Galaxy, more than sufficient to match the
cooling observed in the CO lines.

\end{abstract}


\section{Introduction}

NGC~253 is a nearby ($d \sim 2.5$~Mpc; \citealp{mau96} ), nearly edge-on
($i~\approx~78^{\circ}$, \citealt{pen81}), isolated starburst galaxy with the starburst
strongly concentrated at the nucleus.  Based on the distribution of the 10--30~$\mu$m
continuum, the bulk of NGC~253's starburst is confined to a $\sim 60$~pc region centered
just south-west of the dynamical nucleus \citep{tel93}.  The 1--300~$\mu$m luminosity
within about 30\as\ is $\rm 1.6 \times 10^{10}~L_{\odot}$ \citep{tel80} while the total IR
luminosity of NGC~253 detected by IRAS is $\rm L_{\rm IR} \approx 2 \times
10^{10}~L_{\odot}$ \citep{ric88}.  There is a clear bar in the 2MASS image of the galaxy,
and the kinematics show evidence for orbits in a bar potential (\citealt{sco85};
\citealt[and references therein]{das01}).  The bar likely plays a role in transporting gas
to the nucleus and thus supporting the starburst.  Current estimates of the neutral gas
mass in the central $20'' - 50''$ range from $\rm 2.5 \times 10^{7}~M_{\odot}$
\citep[hereafter HHR99]{hhr99} to $\rm 4.8\times 10^{8}~M_{\odot}$ \citep{hou97}.

As is the case in many of the nearby starburst galaxies, much of the molecular gas in
NGC~253 appears to be highly excited \citep{wil92,mao00,war02}.  Observations of
\jfour\ and \jsix\ transitions of CO \citep{isr02,har91}, as well as \jone\ and \jthree\
transitions of HCN \citep{pag95,pag97} suggest that a large fraction of the molecular gas
in the nucleus of NGC 253 has $\rm n_{H_2} > 10^4 cm^{-3}$ and $\rm T\sim 100 K$ or
higher.  Observations of higher-J CO transitions are critical to measure this warm
molecular gas component, and probe its physical conditions.  In our Galaxy, the warm,
dense molecular gas conditions which excite the mid-J CO lines (J~=~4--8) are typically
associated with molecular cloud surfaces exposed to stellar far-UV radiation.  In these
photodissociation regions (PDRs), the UV dissociates and ionizes the gas as it enters the
cloud and is absorbed by dust.  The result is a layered geometry based on the extinction of UV;
the progression from the exterior inward is ionized, neutral atomic, warm molecular, and
finally cool molecular gas on the interior \citep{th85,wol90,kau99}.  Several of these
regions have been studied in star formation sites within the Galaxy
\citep{har85,sch89,sta93}, and they have been discussed in the context of starburst nuclei
\citep{mao00,car94}.  Other potential molecular gas heating sources are outflows and
cloud-cloud collisions in which shocks convert dynamical energy to thermal energy of the
gas, such as in the circumnuclear disk of the Galactic Center, and in the massive star
formation sites W~51 and DR~21 \citep{jhg87,jaf89}.  Finally, it has been proposed that
much of the molecular gas heating in starburst nuclei may be due to cosmic rays
\citep[hereafter SAH]{suc93}. To better understand the energy balance of the molecular gas
in the starburst nucleus of NGC 253, we have observed the CO \jseven\ transition with our
submillimeter imaging spectrometer, SPIFI on the JCMT.

\section{Observations}
NGC~253 was observed in April 1999 during the commissioning run of the South Pole imaging
Fabry-Perot interferometer (SPIFI) \citep{bra02} at the James Clerk Maxwell Telescope
(JCMT) on Mauna Kea, Hawaii.  The Fabry-Perot interferometer (FPI) was set to a velocity
resolution of 70~km~s$^{-1}$ and was scanned over 620~km~s$^{-1}$ with the center
wavelength of the scan set to the redshifted (250~km~s$^{-1}$) CO \jseven\ rotational
transition.  The FPI scan length was measured with interference fringes of a He-Ne laser,
and absolute frequency reference was an absorption spectrum of the $\rm J\eqq6\too7$ CO
transition from gas in a cell against a hot load.  On this observing campaign the
bolometer array was equipped with 12 functioning detectors arranged on a $4\times4$ grid
with four blanks.  Because of a figuring error in the polyethylene lenses used as a
f-converter, the main beam as measured with Mars had a size of $11.5''$ (FWHM), 1.6 times
bigger than the diffraction-limited beam size of $7''$ per pixel.  Mars was also used for
flux calibration assuming a brightness temperature of 201~K \citep{wri76,rud87}.
Flat-fielding of the array was accomplished by rastering Mars around the array.  For the
period of the observations, the zenith atmospheric transmission was between 0.22 and 0.29,
giving a transmission to the source between 0.05--0.12.  This low and variable atmospheric
transmission provided the greatest calibration uncertainty, which we estimate to be about
30\%.  For the observation of NGC~253 the array was centered at the IR nucleus at
$R.A.=0^h45^m06^s.0$, $Dec.=-25^{\circ}33'40''$ (1950) ((0,0)-position).  We estimate our
pointing accuracy to be 5\as.  The data were obtained by chopping and nodding the
telescope and the data presented here were obtained with 20 minutes on-source integration.
The 50 minute duration of the observations resulted in a field rotation of 18$^\circ$\
about the center of the array, producing a total travel of 7\as\ ($\pm$3.5\as\ about the
mean) for the corners of the array.  Because this is comparable to the pointing
uncertainty, and is substantial only for the corner pixels, we neglect the rotation
and adopt the mean position angle.

\section{Results}\label{sec:results}

\subsection{Comparison with Other Molecular Tracers}\label{sec:comp-with-other}
\begin{figure} 
 \plotone{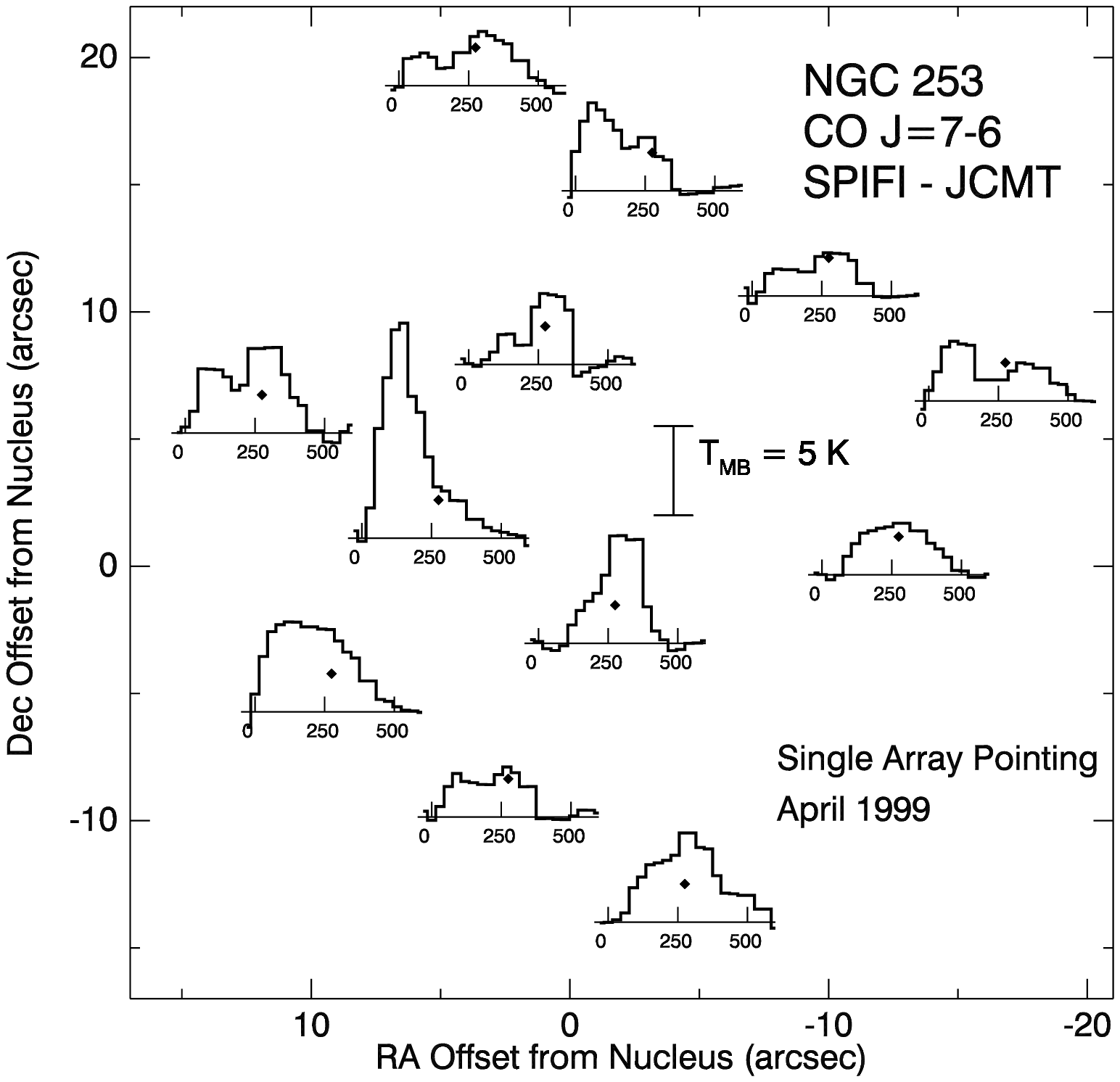}
 \caption{ 
  CO \jseven\ spectra toward the nucleus of NGC 253.
  The (0,0)-position is at the IR nucleus at, 
  $R.A=0^{\rm h}~45^{\rm m}~06^{\rm s}.0, 
  Dec.= -25^{\circ}~33' 40''$ (1950) and the beam size is $11''$.
   \label{fig:point253}}
\end{figure}

Figure~\ref{fig:point253} shows the spectra obtained in this single array pointing toward
the nucleus of NGC~253.  
The data are plotted in main beam
brightness temperature units, correcting for beam efficiency, atmospheric transmission, and
source coupling.  Typical spectra peak at 3--5~K, with a single spectrum showing a peak of
12~K; integrated intensities are between 400 and 1200~K~km~s$^{-1}$.  The integrated
intensities in K~km~s$^{-1}$ are roughly half of the CO \jsix\ intensities of
\citet{har91}, when the \jsix\ antenna temperatures are corrected to a $10''$ main beam.  The
peak temperature of 12 K at $v = 150$~km~s$^{-1}$ in the spectrum just northeast of the
nucleus has a similar peak above the Gaussian profile in the \jsix\
spectrum toward the nucleus.  In addition to the
bright spectra near the nucleus, we detect emission at 30--40\% of the nuclear value at
positions 10\as--20\as\ offset from the nucleus. 
\citeauthor{har91} also detect strong CO
\jsix\ emission at offset positions.  They measure an integrated intensity 60\%
of the nuclear value as far as $14''$ offset, consistent with our observations.

Interferometric observations of millimeter wave CS \citep{pen96}, CO \citep{cms88,car90},
and SiO \citep{gar00} transitions show evidence for a nuclear bar with major axis oriented
70$^{\circ}$ east of north.  The velocity field of this bar shows a
100\kms\ shift from the northeast to the southwest.  Though the \jseven\ map does not have
the resolution or sampling to spatially resolve the bar, it also shows the velocity
shifting from 150-200~\kms\ in the northeast to 250-300~\kms\ in the southwest, showing the
overall kinematics of gas in the bar, in good agreement with the interferometric
observations.  The observed line widths are between 200 and 300\kms, consistent with the 
150 to 240~\kms\ line widths observed in the lower-J CO lines with 12 to 15\as\ beams
(\citet{mau96}; HHR99), convolved with our 70~\kms\ Lorentzian instrumental profile. 

\subsection{Model fit to the nuclear excitation conditions}\label{sec:model-fit}

To make quantitative estimates of the physical parameters of the nuclear molecular gas we
have compiled all CO rotational transitions observed toward the central 15\as\ available
in the literature to compare with predictions of an escape probability, large velocity
gradient model for CO line excitation.  In contrast with previous studies of the molecular
material in starburst galaxies which invoke two components (HHR99) or complex
PDR models \citep{mao00}, {\it our modeling suggests that the observed $\rm ^{12}CO$ and
$\rm ^{13}CO$ intensities can be reproduced with the bulk of the gas mass at fairly high
excitation.}  The line ratios can be fit with a range in conditions from $\rm T = 100\,K$,
$\rm n_{H_2}\sim 5\e{4}\,cm^{-3}$ to $\rm T > 300\,K$, $\rm n_{H_2}\sim2.0\e{4}\,
cm^{-3}$.  Recent measurement of rotational \hh\ lines \citep{rig02} suggests that the
temperatures may be on the low end of the allowed range, and we thus we favor $\rm
T\sim100\, K$ and $\rm n\sim 5\e4\, cm^{-3}$.  Combining the $^{13}$CO transitions
constrains the column density to be only slightly larger than that measured by HHR99 in a
23\as\ beam.  To reproduce the observed intensities with the LVG formalism requires large
velocity gradients ($\rm dv/dz \sim 10\rightarrow80\, km\,s^{-1}\,pc^{-1}$), which are
much higher than is traditionally used, but nevertheless plausible for the starburst
nucleus.  Here we describe our modeling approach and the basis for these conclusions.

Including our results, $^{12}$CO has been observed toward the nucleus of NCG~253 in all
transitions up to \jseven\ except for \jfive, with a variety of beam sizes.  Also, the
optically thin isotopes $^{13}$CO and C$^{18}$O have been observed in transitions up to
\jthree.  Table~\ref{tab:lines} summarizes the $^{12}$CO and $^{13}$CO observations and
the corrections to refer them to the 15\as\ beam.  Measurements for all \jthree\ and lower
transitions are from HHR99.  Coupling corrections, where necessary, are based on power
laws.  The \jtwo\ lines have been observed with both 12\as\ and 23\as\ beams, and the
resulting power law is $\rm I\propto \theta^{-0.76}$.  For \jone, there are no single dish
observations with a beam smaller than 23\as, but based on the centrally peaked \jthree\ /
\jtwo\ ratio map, the \jone\ emission is probably more extended than \jtwo.  We adopt a
power law of $\rm I\propto \theta^{-0.4}$, but estimate a 20\% uncertainty in the derived
intensity, which is sufficient to accommodate a power law exponent between 0 and 0.76.  For
\jthree, the intensities are measured from the published spectra in 15\as\ beams.  For
\jfour\, there are published measurements with both a 10.6\as\ beam \citep{isr95} and a
22\as\ beam \citep{isr02}.  Allowing these two measurements to define a power law gives
$\rm I\propto \theta^{-1.02}$, and we use this to determine the intensity within 15\as.
The \jsix\ measurement is from \citet{har91} in $\rm T_R^{*}$ units, and is corrected to
15\as\ according to their stated 8\as/30\as\ composite beam profile.  The \jseven\
observation from this work is referred to an 11.5\as\ main beam based on observations of
Mars, with a small correction for the 15.8\as\ planetary disk.  The estimated
uncertainties are also included in Table~\ref{tab:lines}.  The beam filling corrections
become the dominant source of uncertainty in the \jone\ (described above) and \jtwo\
lines, we estimate this to be about 15\%.  For the \jthree\ transition in a 15\as\ beam,
we estimate 20\% for the overall uncertainty including systematics and statistics.  For
the \jfour\ and higher submillimeter transitions, the uncertainties in the measurements
themselves are substantial, as much as 30\%.

The LVG model calculates intensities of rotational CO lines
given a molecular hydrogen density, a temperature, and a velocity gradient.  The
calculation is similar to that used in other multi-line analyses, iteratively solving for
the level populations up to $\rm J\eqq15$.  We use the escape probability $\beta =
\left(1-e^{-\tau}\right)/\tau$ of \citet{gk74} appropriate for spherical clouds undergoing
gravitational collapse.  \hh~--~CO collisional excitation rates are those of \citet{fl01},
increased by 21\% to account for collisions with He \citep{mck82,vc88,fl85,fl01,sch85}.

\begin{deluxetable}{cccccccc}\tablewidth{0pt}
\tablecaption{CO observations toward the NGC 253 nucleus\label{tab:lines}}
\tablehead{\colhead{Transition}&\colhead{$\nu$
[GHz]}&\colhead{Obs. Beamsize}&\colhead{$\rm \int{T_{MB}\,dv}$}
&\colhead{Corr'n to 15\as}&\colhead{Int. in 15\as$^1$}&\colhead{$\pm$ \%}}
\startdata
$^{12}$CO \jone & 115.3 & 23\as & 920 & 1.2 & $1.72\e{-6}$&20\\
$^{12}$CO \jtwo & 230.5 & 12\as & 1060 & 0.84 & $1.37\e{-5}$&15\\
$^{12}$CO \jthree &345.8 & 15\as & 998 & 1 & $6.5\e{-5}$&20\\
$^{12}$CO \jfour &461.4 & 22\as, 10.6\as&1019, 2150&1.48, 0.70&$1.50\e{-4}$&30\\
$^{12}$CO \jsix &691.5& 8\as\ / 30\as &861&1.59 &$4.6\e{-4}$&30\\
$^{12}$CO \jseven&806.7& 11.5\as\ / 60\as &1370&0.677&$5.0\e{-4}$&30\vspace{0.1in}\\ 

$^{13}$CO \jone & 110.2 & 23\as&80 &1.2&$1.32\e{-7}$&20\\ 
$^{13}$CO \jtwo & 220.4 & 12\as&134 &0.84&$1.24\e{-6}$&15\\
$^{13}$CO \jthree & 330.6 & 15\as&145&1&$5.4\e{-6}$&20\vspace{0.1in}\\  
\enddata

\tablecomments{$^1$ Integrated intensity in $\rm erg\,s^{-1}\,cm^{-2}\,sr^{-1}$.  The
intensities for \jthree\ and lower are taken from \citet{hhr99}. \jfour\ is from 
\citet{isr95,isr02}, \jsix\ from \citet{har91}, and \jseven\ from this work.  The couplings and
uncertainties are described in the text.}
\end{deluxetable}

\subsubsection{$^{12}$CO Intensity Ratios}\label{sec:12co-intens-rati}
 
A sensible approach for comparing an LVG model with observations is to use the line
intensity ratios to determine the physical parameters (temperature, density, and opacity)
while the absolute line intensities determine the total column density.  In addition to
the $^{12}$CO intensity ratios, the optically-thin isotopes can be useful in constraining
the $^{12}$CO optical depths, which are greater than unity for the lines of interest.  In
order to concentrate on the nuclear gas, we do not put much emphasis on the \jone\
transition.  Since it connects the ground state and and the $\rm J=1$ level is only 6 K
above ground, the \jone\ transition traces any molecular material in the beam.  A
related and more confusing problem is the potential for self-absorption by quiescent
material along the line of sight but exterior to the nucleus, a particular concern for
nearly edge-on systems such as NGC~253.  This may be the reason
that any model which fits the \jtwo\ and higher transitions over predict \jone.  In an
approach similar to that of HHR99, we invoke a small diffuse component which is negligible
energetically, but which emits primarily in \jone\ to account for its observed intensity.

For a given LVG calculation, the parameters which determine the line intensity ratios are
the temperature, density, and column density of CO per unit velocity interval, \ncodv,
called the opacity parameter.  \ncodv\ is a combination of density, velocity gradient of
the gas, and CO abundance, $\rm N_{CO} / dv = X_{CO} n_{H_2} (dv / dz)^{-1}$ where $\rm
dv/dz$ is the velocity gradient in $\rm km/s\, pc^{-1}$.  We have assumed a CO abundance
\xco\ of $8\e{-5}$, the value measured for galactic clouds \citep{fre82}.  While the
metallicity in the nucleus of NGC~253 is as much as 3 $\times$ solar \citep{mau96}, the large observed [CI]
intensities suggest that, relative to the Galaxy, the gas-phase carbon budget includes
a much larger neutral carbon fraction, and smaller CO fraction \citep{isr02}, thus the CO
abundance is not clearly much higher in spite of the large metallicity.  Intensity ratio
calculations for a higher CO abundance are identical to those with the canonical \xco\
value, but with the velocity gradient decreased by the same ratio.  Raising the CO
abundance or lowering the velocity gradient increases the the optical depth in the lines.
Figure~\ref{fig:ntplot} plots the results of LVG calculations of diagnostic line ratios as
a function of gas temperature, density, and velocity gradient.  We have found that the
\jseven\ to \jthree\ intensity ratio is a good measure of the overall excitation, and that
matching this ratio typically satisfies all the lines above \jone\ to within the
observational uncertainties.
%

As can be seen from the solid contours in Figure~\ref{fig:ntplot}, the $^{12}$CO \jseven\
to \jthree\ intensity ratio of 7.7 means that the gas is of high excitation.  Because the
run of line intensity with J is increasing up to \jseven, there is no unique solution, nor
even upper limits to the temperature or density.  Lower limits, however, are strong.  For
temperatures less than 300~K, the density must be greater than $\rm 2-3\times 10^{4}\,
cm^{-3}$.  A temperature less than 150 K requires a density greater than $\rm \sim
4\times10^{4}\, cm^{-3}$, clearly different from the typical Galactic molecular cloud.  To
first order, these results are independent of the velocity gradient and CO abundance
because the lines are optically thick, even at velocity gradients as large as 80 \kmspc.

\subsubsection{$\rm ^{13}CO$ Intensities}\label{sec:13co-isotope}

\begin{figure}
  \plotone{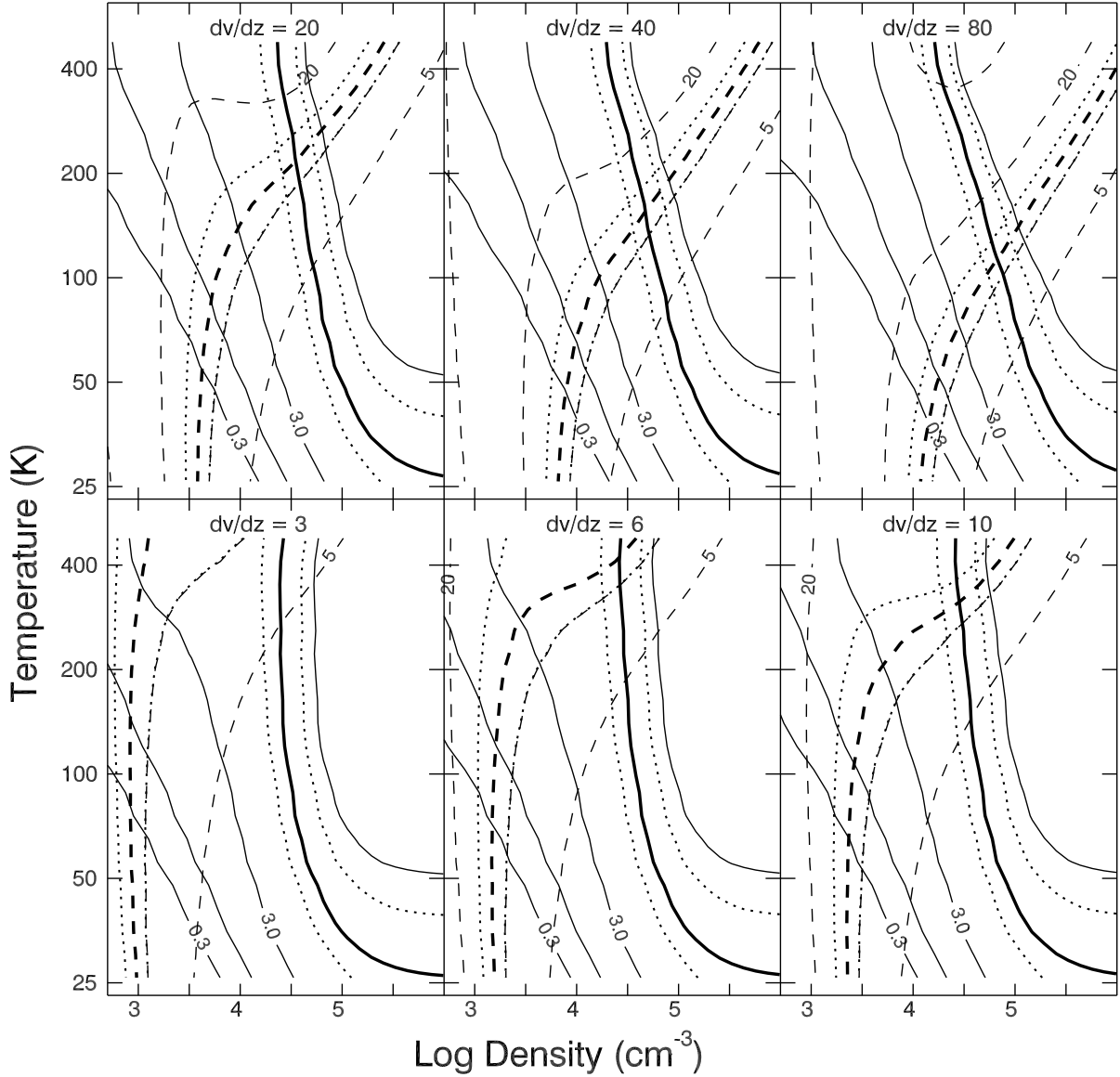} 
\caption{LVG model calculations compared with observed line
   ratios.  Contours of the $^{12}$CO \jseven\ to \jthree\ intensity ratio are plotted as
   solid lines (values of 0.3, 1, 3, 10).  Calculations are also performed for the
   $^{13}$CO transitions assuming an abundance ratio of 50.  The $^{12}$CO to $^{13}$CO
   \jthree\ ratio is shown with the dashed lines (values of 5, 10, 20, 40).  For both ratios,
   the observed values are shown with thick lines, bracketed with thin dotted lines
   showing 20\% uncertainty.  Scaling the CO abundance from the canonical $8\e{-5}$ is
   equivalent to adjusting the velocity gradient in the opposite
   direction with the same factor.  With the canonical CO abundance, the velocity gradient
   must be greater than 10\kmspc\ to reproduce the data.   Larger CO abundances require even
   larger velocity gradients.  \label{fig:ntplot}}
\end{figure}

If their abundances relative to the main isotope are known, the intensities in the
$^{13}$CO transitions can be used to determine the $^{12}$CO optical depths.  This is the
approach taken by HHR99 in their study of the nuclear regions of NGC~253 with the lower-J
lines.  The key to their analysis is the assumption of a $^{12}$CO to $^{13}$CO abundance
ratio of 40--50, originating from analysis by \citet{hen93,hm93}.  Our results are
consistent with a ratio on the upper end of this range, and we have chosen 50 as a value
to use in the model plots.  Included in Figure~\ref{fig:ntplot} as dashed lines are the
$^{12}$CO to $^{13}$CO \jthree\ intensity ratios predicted by the LVG models.  The
observed ratio of 12 is plotted as a thick dashed line.  To correctly obtain the line
intensity ratios between the two isotopes requires running the full LVG model for the
optically thin species, since radiative trapping is important in determining the level
populations of $^{12}$CO but is much less so for the optically thin species.  Because
the $^{13}$CO transitions are optically thin for all relevant conditions, the values
plotted in Figure~\ref{fig:ntplot} can be shifted to accommodate a different isotope
ratio.  In the density-temperature plane, a larger isotope ratio is considered by choosing
a line ratio contour with a value decreased from the observed 12 by the same factor.


In the panels of Figure~\ref{fig:ntplot}, the temperature and density at which the two
observed line ratios meet is the regime of possible solutions.  It is evident that small
velocity gradients are unable to generate the observed line ratios.  At these small
velocity gradients, the conditions which excite upper-J lines generate large optical depths
in the $^{12}$CO lines, and their fluxes are too small compared with the optically thin
isotope.  Increasing the velocity gradient does provide possible solutions, again, with
high excitation conditions in both temperature (T$>$100 K) and density ($\rm
n_{H_2}>4\e5$).  Though unconventional in LVG analyses, the large velocity gradient is
plausible for the starburst nucleus, and is discussed in Section~\ref{sec:velocity-dispersion}.


\subsubsection{CO Column Density}\label{sec:co-column-density}

\begin{figure}
  \plotone{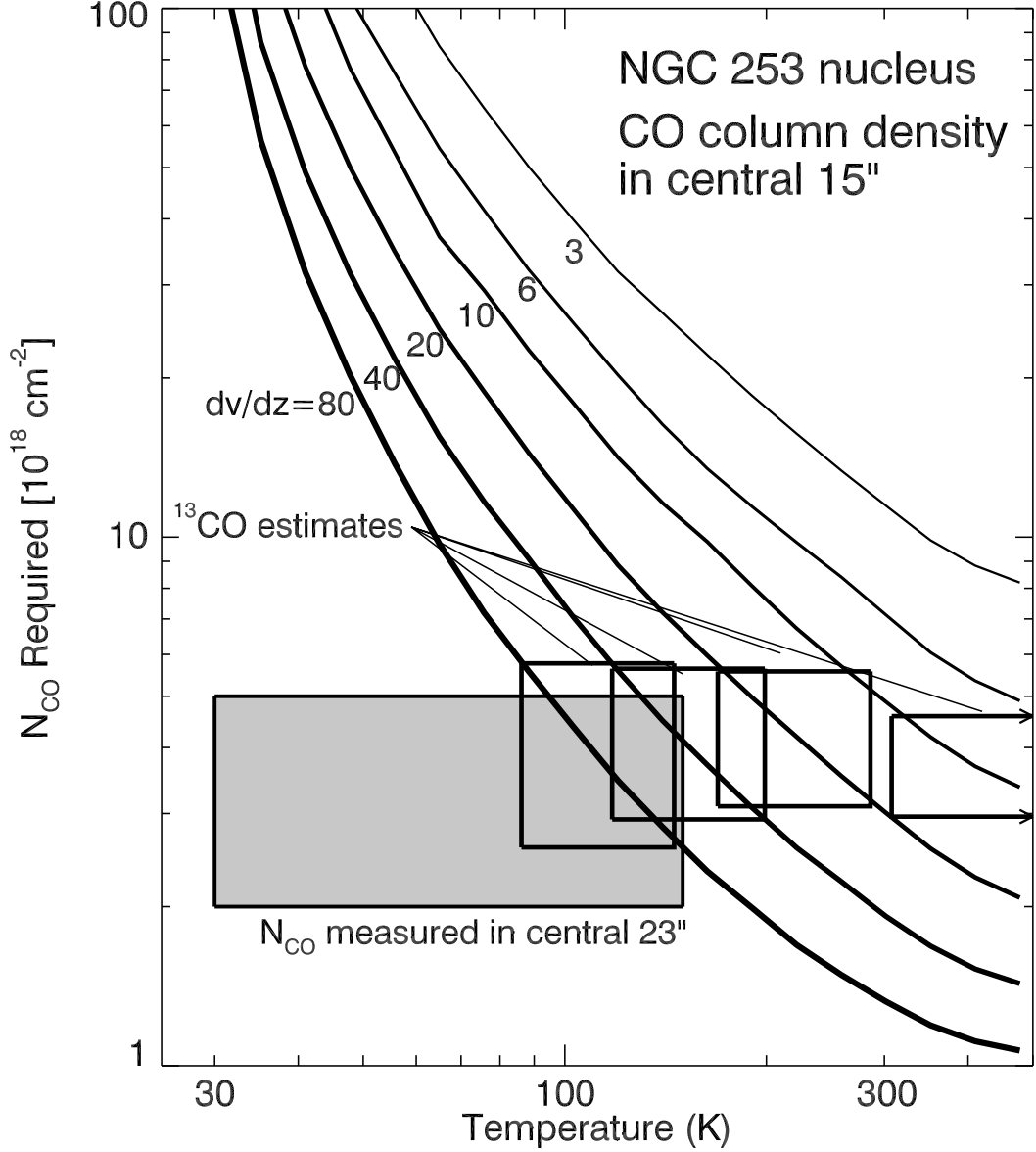}
    \caption{Relationship between temperature, velocity gradient, and column density
    required to reproduce the observed intensities.  At each point, the density is determined to match the
    $^{12}$CO \jseven\ to \jthree\ line ratio, then the observed integrated intensities are compared
    with the LVG model to determine the column density required.  Higher temperatures and
    lower optical depths require smaller column densities.  The constraints based on
    the $^{13}$CO to $^{12}$CO \jthree\ ratio are shown as boxes for \dvdz = 10, 20,
    40, and 80 \kmspc.  The sizes of the boxes are estimates of the uncertainties. \label{fig:cdplot}}
\end{figure}

Even though the $^{12}$CO lines are optically thick, their observed intensities put lower
limits on the CO column density independent of the optically thin isotopes.  The intensity
in a line is determined by the product of column density in the upper level and the
optical depth factor $1-e^{-\tau}$, hence a lower optical depth (larger velocity gradient)
generates more emission for a given CO column density.  The curves in
Figure~\ref{fig:cdplot} show the column density required by the LVG model to reproduce the
observed intensities.  For each velocity gradient, the curve is parametrized in terms of
temperature, with the density determined according to the observed $^{12}$CO line ratio shown in
Figure~\ref{fig:ntplot}.   
Examination of Figure~\ref{fig:cdplot} indicates that the column density of warm gas must be substantial.
For example, if the gas has a temperature below 200~K and velocity gradient below 40
\kmspc, the column density must be larger than $3\e{18}\rm\,cm^{-2}$, a lower limit which
is comparable to the $3.5\pm 1.5\e{18}\rm\,cm^{-2}$ estimated in the central 23\as\ with
the \jthree\ and lower transitions (HHR99).



To improve upon the lower limits provided by the $^{12}$CO transitions alone, we incorporate
the $^{13}$CO intensities.  The boxes in Figure~\ref{fig:cdplot} show the column densities
required by the LVG model under the conditions corresponding to the intersections of
the two diagnostic line ratio curves in the panels of Figure~\ref{fig:ntplot}.
Boxes are plotted for values of \dvdz\ = 10, 20, 40 and 80 \kmspc, as velocity gradients
less than 10 \kmspc\ do not allow reasonable solutions.  The sizes of the boxes correspond
to a $\pm$30\% uncertainty in temperature, and an corresponding uncertainty in column
density given by local slope of the T-$\rm N_{CO}$ relation.  The locus of these regions
indicates that the CO column density is between 3--7$\rm\e{18}\,cm^{-2}$, for all allowed
values of the excitation.  This is consistent with the $^{12}$CO upper limits, and is
comparable to the value derived by HHR99 for the central 23\as.
The results of our LVG analysis can be summarized as follows: the molecular gas in the central
180 pc of NGC~253 is much more highly excited than that traced with only the lower-J lines
in a larger 280 pc beam.  The gas is warmer than 100~K, and has a density greater than $\rm
2\e{4}\, cm^{-3}$.  The $^{12}$CO lines have optical depths not larger than 2--4,
requiring velocity gradients $>10$ \kmspc, larger than traditionally employed in LVG
studies.  Even larger velocity gradients ($>20$ \kmspc) are required if the temperature is constrained to
be below 200~K.  The CO column density is well-determined by
the $^{13}$CO lines: $\rm N_{CO} = $3--7$\e{18}\rm \,cm^{-2}$, only slightly larger than
the values estimated in a larger beam.  Finally, while the $^{13}$CO
\jone\ transition requires an additional component of cool and/or diffuse gas, this
component is small by mass, negligible energetically, and may not even be associated with
the nuclear region.  There is no reason to invoke a large component of cool gas
as would be required, for example, in a study of the central regions of the Milky Way.


\subsubsection{The Large Velocity Gradient}\label{sec:velocity-dispersion}

Previous LVG studies of NGC~253 and M82 have employed a velocity gradient of around
1~\kmspc, which is approximately the ratio of the total observed velocity width to the
size of the observed region (HHR99; \citet{mao00}), and comparable to the values used in
studies of Galactic molecular clouds.  This is a lower limit to the true velocity gradient
in the material, appropriate if it were only subject to the central potential of the
galaxy with no local motion.  In reality, clouds and clumps will be subject to their own
gravity as well as that of the stars around them.  The escape velocity of a self
gravitating cloud of radius r, and density $\rho$ determines a velocity gradient given by
$\rm dv/dr\sim \sqrt{8\pi/3 G \rho}$, which gives 10 \kmspc, for a density of $\rm
n_{H_2}\sim 4.5\e4 \ccm$.  Models on the low end of the allowed
temperature range, e.g. T~$\sim$~120 K require a velocity gradient of order 40~\kmspc, a
factor of 4 larger than this escape velocity.  This is not implausible; if the
star-forming clouds in NGC~253 are similar to the Orion molecular cloud core, then one
might expect the stellar mass density to be about 4 times the gas density \citep{od01},
which would double the effective escape velocity.  Furthermore, \citet{car94} find
the pressure of the ionized gas in the central 45" of NGC~253 is $\rm \sim 7\e6\, K
\,cm^{-3}$ - larger than the thermal pressure within our modeled molecular gas.  This high
pressure ionized gas has a large volume filling factor (0.01 to 0.3) within their much
larger (45\as) beams.  It is thus quite plausible that within our 15\as\ modeled region the
ionized gas place a important role in cloud confinement, further increasing the velocity
dispersion in the gas.  
In short, the large velocity gradients required for a good model fit are very plausible
given the dynamics in the nucleus of NGC~253.   

\subsection{Comparison with rotational \hh\ lines}\label{sec:comp-with-rotat}

Because the \jseven\ transition is not high enough to see the turn-over in CO excitation,
the CO analysis does not provide a good upper limit to the gas temperature.  Fortunately,
we can use the ISO \hh\ pure rotational line observations to provide an important
temperature constraint.  \citet{rig02} have recently reported ISO SWS observations of the
pure rotational lines of \hh\ from the central 14\as~\by~20\as\ regions of NGC~253.  These
quadrupole \hh\ lines have low radiative rates, so they are easily thermalized and
optically thin.  In principle the lines can be used to measure temperature and column
density.  However, even the lowest \hh\ rotational transition energy is 515~K, thus the
observed line emission is heavily biased toward the very warmest regions (which may be
negligible by mass), and each observed transition is an exponential function of gas
temperature.  Not surprisingly, the line intensities are not consistent with a single
excitation temperature.  Based on the S(0), S(1), S(2), S(5), and S(7) lines,
\citet{rig02} conclude that there must be a substantial mass of warm, dense gas in the
nucleus of NGC~253, broadly consistent with our results.  More specifically, the modeled
molecular hydrogen column density of $\rm N_{H_2} = N_{CO}/ X_{CO}\sim $4--8$\e{22}\rm\,
cm^{-2}$ is consistent with the S(0) data presented by \citet{rig02} for gas temperatures
in the range of 95--120~K.  We therefore favor models on the low end of the temperature
range allowed by CO analysis.  Because of the exponential dependence on temperature, this
\hh\ temperature constraint is a rather weak function of \hh\ column density, hence also a weak
function of CO column density and CO abundance.

\begin{figure}
 \plotone{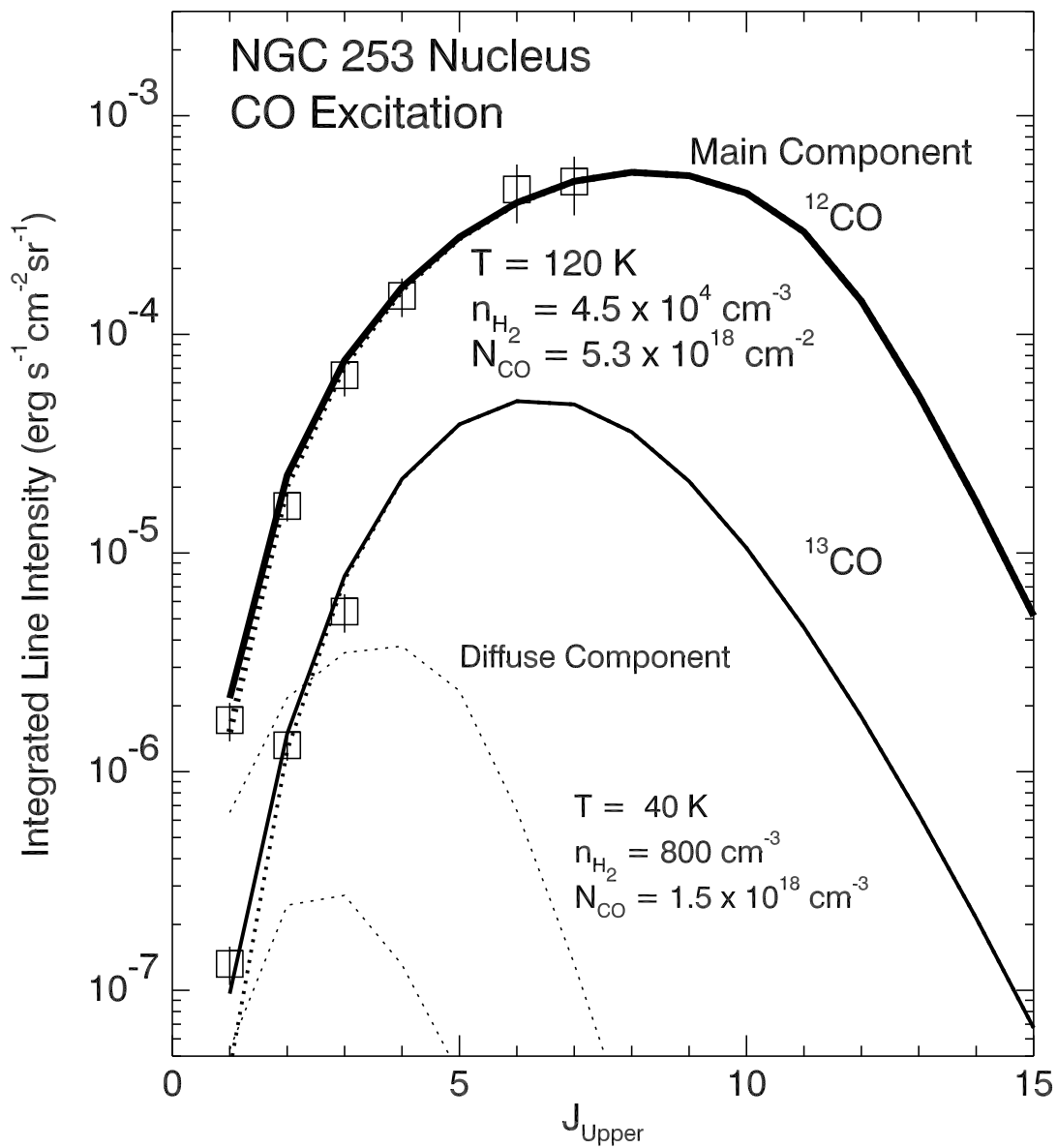} 
\caption{Integrated $^{12}$CO and $^{13}$CO line intensities
 in the nucleus of NGC 253 with our adopted model of the excitation.  The data are compiled from
 the literature and smoothed or adjusted to correspond to a 15\as\ beam, see
 Table~\ref{tab:lines}.   
The solid lines are the total, the thin dotted lines in the lower left show a diffuse
component which is energetically negligible except for a contribution to the \jone\ lines.
\label{fig:fit253}}
\end{figure}

We combine this \hh\ result with the parameter ranges allowed by the LVG
analysis, and the physical constraint of a small velocity gradient to arrive at our
preferred model, shown in Figure~\ref{fig:fit253}.  This model uses \dvdz\ = 40 \kmspc,
toward the bottom of the range allowed by the low temperature and the observed $^{13}$CO
intensity.  As described above, the lower velocity gradients are more consistent with
gravitationally bounds clouds.  The modeled conditions are $\rm T=120\, K$, $\rm n_{H_2} =
4.5\e{5}\,cm^{-3}$, and the CO column density required to generate the observed line
intensities is $\rm 5.3\e{18}\, cm^{-2}$.  At this temperature, about 40$\%$ of the
observed \hh\ S(1) flux is generated in the bulk component, and the contributions to the
higher-J \hh\ lines (which require excitation of levels with $\rm E~>~1500~K$) are less
than 10\% of the observed line fluxes.  As an estimate of the uncertainties, we note that
even with the temperature constrained, a larger velocity gradient still allows a smaller
column density.  Decreasing the velocity gradient, or increasing the CO abundance can
permit a larger column density, but this quickly becomes inconsistent with the $^{13}$CO
observations unless the $^{13}$CO relative abundance is decreased below our adopted 1/50.
Combining all the uncertainties in the abundance ratio, modeling, and observations, we
estimate an uncertainty of 40\% for the final adoped column density: $\rm N_{CO}\sim
3.5-7\e{18}\,cm^{-2}$.  Our model parameters are summarized in Table~\ref{tab:props}.

\subsection{Warm Gas Mass}\label{sec:warm-gas-mass-1}
With the assumed CO abundance relative to \hh\ of $8\e{-5}$, the CO column density
corresponds to a mass of excited molecular gas in the central 15\as\ of
2.5--5.1$\e{7}\,\rm M_{\odot}$.  In reality, the CO abundance is not well-known.  The
abundance may be larger than the canonical Galactic value, given the 3$\times$ solar
metallicity (but see discussion above), and this would result in a smaller gas mass by as
much as a factor of 3.  Based on an assumed carbon gas to dust ratio and the measured
fractions of gaseous carbon bearing species, HHR99 infer a CO abundance relative to \hh\
of $2.2\e{-4}$, that is, about the factor of 3 larger than our adopted value.  With this
abundance, the gas mass would be smaller: 0.9--1.8$\e{7}\,\rm M_{\odot}$.  We do note a
larger CO abundance demands even larger velocity gradients (more than 30 \kmspc) to allow
working solutions.  For comparison, the most recent estimate of the gas mass based on 1.3
mm dust continuum is $4.4\e7\,\rm M_{\odot}$ (HHR99, \citet{kru90}).  Since the 1.3 mm
continuum has a half-power size of only 16\as\ $\times$ 11\as\, much of this flux arises
within our within our 15\as\ beam, and traces the same material.  The agreement with our
derived gas mass suggests that while our CO abundance may be slightly high, it is not
off by more than a factor of 2.

\subsection{Suitability of Multiple-Component Models}\label{sec:suit-two-comp}

It is traditionally thought that mid-J CO emission originates in the PDR surfaces of
molecular clouds exposed to UV starlight.  In this scenario, the warm excited gas is
separate from the bulk of the molecular gas, and usually represents a small fraction of
the total mass.  Observations of several PDR regions in the Galaxy have motivated modeling
which includes detailed PDR chemistry and radiative transfer to predict the line emission.
\citet{kos94} have constructed detailed models of PDRs on the surfaces of small clumps to
predict $^{12}$CO and $^{13}$CO line intensities in star-formation regions.  Relative to a
simple plane-parallel model, the clumpy geometry allows an increase in local gas density,
and allows UV photons to penetrate to a greater fraction of the gas, increasing the
overall molecular excitation level.  While such models are successful in reproducing the
intensity ratios between $^{12}$CO lines, they fail to account for both the large
integrated intensities and the large $^{12}$CO to $^{13}$CO intensity ratios observed in a
starburst nucleus such as NGC~253.  For example, \citet{kos94} model slabs with thickness
given by $\rm A_V = 10$, illuminated from both sides, including chemistry of both
$^{12}$C- and $^{13}$C-bearing species.  The ratios among the $^{12}$CO lines are well-fit
with $\rm n_{H_2}=10^5\, cm^{-3}$, $\rm \chi_{UV} = 10^5$ Habing ($\rm=1.6\e{2}\,
erg\,s^{-1}\,cm^{-2}$), but to generate the observed \jseven\ flux in the central 15\as\
of NGC~253 with a superposition of these regions would require between 5 and 20 times the
CO column density estimated with our LVG analysis.  The $^{13}$CO lines show the same effect --
the predicted \jthree\ $^{12}$CO to $^{13}$CO intensity ratio is around 3, compared with
the observed value in 15\as\ of $\sim 12$ (HHR99).  Thus the existing PDR models
are not appropriate for the global properties of a starburst nucleus.  The key problem is
the high optical depth in the PDR models, resulting from the small assumed velocity width per
gas column.  The effective velocity gradient of the clumpy PDR model is around 9~\kmspc,
too low as we have discovered in Section~\ref{sec:model-fit}.  

To consider a PDR scenario with the appropriate optical depths in the lines for NGC~253,
we have examined LVG models with two distinct gas components.  Though this approach is
not as rigorous as a full treatment of the chemistry and UV penetration used by the
Galactic PDR models, to first order it accounts for the combination of warm
UV-exposed surfaces and cool cloud interiors which together would generate the observed CO
emission. For NGC~253, we have found that while it is possible to fit the observed
intensities without the full mass of gas being highly excited, the warm fraction must be
substantial.  To reproduce the observed \jsix\ and \jseven\ transitions, at least 30\% of
the total column density estimated with the $^{13}$CO ratio must have high excitation
conditions.  Furthermore, the curves in Figure~\ref{fig:cdplot} shows that this scenario
requires an extremely large velocity gradient ($\sim$100~\kmspc) and/or very high
temperatures ($\sim$300 K) in this 1/3 warm fraction in order to reproduce the observed intensities.
Moreover, as we outline in Section~\ref{sec:uv-energy-sources}, even with only 1/3 of
the total column density accounting for the high-J line emission, the amount of warm
molecular gas would still be 7--10$\times$ greater than the PDR-associated atomic
gas, so the UV heating is not a viable origin of the observed CO emission.

\section{Discussion}

\subsection{Energetics of the Molecular Gas in NGC 253}\label{sec:energy}

Independent of the details of the excitation and radiative transfer model, the general
shape of the CO intensity vs. J curve measures the luminosity of the molecular gas in the
central 15\as\ of NCG 253.  Taking our favored model plotted in Figure~\ref{fig:fit253} as
a guide, the total flux in the CO lines from the modeled region is $\rm 1.5\e{-11}\,
ergs\,s^{-1}\,cm^{-2}$.  This is 25\% of the [C~{\small II}] 158~$\mu$m and [O~{\small I}]
63~$\mu$m intensities from the entire inner $45''$ \citep{car94}, and a substantial $1.4
\times 10^{-4}$ fraction of the total IR luminosity of the galaxy.  At a distance of 2.5
Mpc, this molecular gas flux corresponds to a luminosity of $\rm 2.9\e{6}\,L_{\odot}$,
giving a luminosity to mass ratio of $\sim 0.1~L_{\odot}/M_{\odot}$ for the gas containing
CO based on the mass estimates of Section~\ref{sec:warm-gas-mass-1}.  The large luminosity
in the CO lines is a simple consequence of the fact that such a large fraction of the gas
mass is highly excited.  In what follows, we examine potential heating sources to generate
the large luminosity.  Simple arguments show that traditional heating sources for warm
molecular gas -- far-UV photons and cloud-cloud collisions -- are unlikely to contribute
substantially.  {\it In contrast, the enhanced cosmic ray flux of the starburst nucleus
can easily match the observed cooling in the CO lines, and is likely responsible for the
high excitation conditions inferred for the bulk of the molecular ISM.}

For completeness we note that the CO lines are a good measure of the total cooling of the
molecular gas.  The five \hh\ lines observed by \citet{rig02} suggest that all the
rotational lines combine to produce a flux of at most $1\e{-11} \rm
ergs\,s^{-1}\,cm^{-2}$, less than the CO lines.  Another potential coolant is C$^0$, which
can have an abundance comparable to that of CO.  However, it only has two transitions
each of which could be comparable to a single CO line if optically thick
and thermalized; thus C$^0$ cannot be a dominant coolant.  Also, we note that though the
modeling described above carries uncertainties as great as 40\% in the derived parameters,
the discussion below typically concerns factors of 3--30, and the general arguments hold
independent of the modeling details.

\begin{deluxetable}{lc}\tablewidth{0pt}
\tablecaption{Dense Molecular Gas Properties in the Central 180 pc of NGC 253\label{tab:props}}
\tablehead{\colhead{Parameter}&\colhead{Value}}
\startdata
\multicolumn{2}{c}{\it Adopted}\\
CO abundance (X$_{\rm CO}$) & $8\e{-5}$ \\
\vspace{0.1in}$^{12}$CO / $^{13}$CO abundance ratio & 50 \\ 
\multicolumn{2}{c}{\it Modeled}\\
N$_{\rm CO,\, beam,\, dense}$ & 3.5--7$\e{18}\,\rm cm^{-2}$ \\
Gas density ($\rm n_{H_2}$) & 3--6$\e{4}\rm\, cm^{-3}$ \\
Temperature ($\rm T$) & 100--140 K \\
Velocity gradient & $>30\rm\, km\,s^{-1}\,pc^{-1}$ \\
\vspace{0.1in}Dense gas mass & 2.5--5$\e{7}\,\rm M_{\odot}$\\ 
\multicolumn{2}{c}{\it Derived with best model}\\
Dense gas filling factor ($\phi_V$)& $2.6\e{-3}$ \\
Cloud radius & 2.0 pc \\
Number of clouds & 330 \\
\enddata
\end{deluxetable}




\subsubsection{Stellar UV Heating}\label{sec:uv-energy-sources}

In Galactic starformation regions, warm CO is typically produced by UV impinging on the
surfaces of molecular clouds.  As it penetrates a cloud, the UV is attenuated, producing
in turn ionized, atomic, and warm molecular layers before it is extinguished \citep{th85}.
The thermal pressure implied by our model fit to the CO lines is $5.4\e6\,\rm K\,cm^{-3}$, similar to the
pressure derived with atomic PDR line ratios of $2.9\e6\,\rm K\,cm^{-3}$ \citep{car94}.
This is consistent with a model in which the PDR gas forms the surface of the warm
molecular clouds.  The penetration of UV into a cloud is determined by the dust opacity,
proportional to the gas column density according to
\begin{equation} \label{eq:av}
 A_{\rm V} = 2 N_{\rm H_{2}} / 1.9 \times 10^{21} {\rm cm}^{-2}. \end{equation} Here we
 have assumed solar dust abundance.  (This is not unreasonable---though the metallicity in
 the nucleus of NGC~253 may be 3$\times$ solar \citep{mau96}, there is evidence for
 substantial destruction of dust grains in an unusually large gas-phase Si abundance
 \citep{car94}).  In standard Galactic PDRs, the first $\sim$ 3 magnitudes from the
 surface of the cloud are photodissociated and cool through the atomic lines, while the
 warm molecular gas exists from $\rm A_V \sim$~3--6.  To first order, the masses of these
 two components are comparable.  Yet in the nucleus of NGC~253, \citet{car94} measure
 $2.4\e6\,$\ms\ of photodissociated gas traced by the 158$\rm\,\mu m$ [C{\small II}] line
 in a large (45\as) beam. Scaling by the observed molecular and far-IR
 continuum morphology \citep{cms88}, \citep{tel93}, we estimate only 56\% of the
 photodissociated gas traced with the large beam falls within our 15\as\ beam.  Therefore,
 the photodissociated gas mass is about a factor of 20--30 smaller than the 2--5$\e7$ \ms\ of
 warm molecular gas.  There is no reasonable geometry in which two layers of comparable
 thickness can have volumes different an order of magnitude.  It is possible
 that the distribution of heating between the molecular and atomic components could be
 made to favor the molecular gas with soft or weak UV fields relative to standard PDR
 templates.  However, the conditions derived with the atomic lines are consistent with
 standard photodissociation region (PDR) models -- the UV field can be understood as
 arising from O~7.5 stars with an effective temperature of 34,500~K \citep{car94,eng98},
 so scenarios invoking soft or weak UV fields are unlikely.

As pointed out in Section~\ref{sec:suit-two-comp}, as little as 30\% of the total CO could
 provide the line luminosities in a warm component, emitting more efficiently lower
 opacity in the lines.  The remaining 70\% of the molecular gas would then be in a cool,
 more diffuse component, possibly the UV-shielded cloud cores.  This scenario is
 attractive because UV-shielded regions are required to host the HCN observed toward the
 nucleus \citep{pag95}, which has a low dissociation energy ($\sim$ 5~eV).  However, a
 factor of 3 reduction in warm CO-containing gas mass would still leave a factor of 7--10 more
 warm molecular gas than atomic gas.  Thus standard PDR models which invoke stellar
 far-UV photons to heat the surface of molecular clouds are not a good match for the
 heating of the bulk of the molecular gas in the nucleus of NGC~253.

\subsubsection{Dynamical Heating}\label{sec:dynam-energy-sourc}

Another potential explanation for the large molecular luminosity is dynamical energy
sources.  By heating molecular gas over large velocity widths, shocks can dramatically
increase the efficiency of CO radiation.  One source of shocks are outflows from young
stars or star associations, examples in our Galaxy include the W~51, W~49 and DR~21
regions \citep{jhg87,jaf89}, in which bright CO \jseven\ emission is observed with
$\sim$20--50~\kms\ linewidths.  In these sources, the CO luminosity in the broad-line
shock-heated outflow components dominates that of the narrow-line PDR component.  Another
clue is the bright SiO observed in the central 150~pc of NGC~253 \citep{gar00}; the large
implied abundance of SiO is attributed to outflows from massive young stars in a manner
similar to the Galactic regions W~51 and DR~21.  However, the comparison cannot apply
globally in the nucleus of NGC 253.  First, the excitation conditions in these Galactic
shock regions are much higher that what we observe for the average properties of the
nucleus of NGC~253.  Observations of CO \jtwo, \jseven\ and \jsixteen\ toward W~51 IRS~2,
for example, indicate that the gas temperature are 500--1000~K, unlikely for the bulk of
the gas in NGC~253 given the rotational \hh\ observations suggesting lower temperatures.
More importantly, in NGC~253, the near-IR \hh\ emission is characteristic of UV fluorescence,
rather than the thermal spectrum produced by shock heating \citep{eng98} found in the
Galactic outflow sources.

Shocks can also be generated with large-scale gas motions: supernova blast waves or spiral
density waves.  The nuclear starburst in NGC~253 is approximately bounded by an inner
molecular ring with diameter 120~pc which corresponds to the inner Inner Linblad Resonance
(iILR) of the galaxy's bar.  The large non-circular motions observed (as much as
40--50~\kms) are attributed to gas accumulation and orbit crowding (see \citet{bc96}), in
the deep potential of the stellar mass, and it is plausible imagine that shocks from cloud
collisions would be common.  However, the difficulty again is that if the majority of the
CO luminosity is to come from cloud collision shocks, they must be slow ($v \sim
10$~km~s$^{-1}$), non-dissociative shocks in order to reproduce the observed CO line
ratios, and not excite the ro-vibrational \hh\ lines with a thermal spectrum.  At these
velocities, the collision timescale and energy available for collisions between clumps
becomes small.  Considering the molecular gas to come from assumed r$\sim2~\rm pc$, $\rm
n_{H_2} = 4.5\e4$ clumps, the mean separation of clumps $\lambda$ is around 25 pc.  With a
relative velocity of 10~\kms\, collisions will occur on timescales of
\begin{equation}
\rm \tau \sim \lambda^{3} / (\sigma v) \sim 1\e8 yr
\end{equation} with an available energy $E = \frac{1}{2} M_{\rm clump} v^2$.  Such
collisions, then, could in principle result in a luminosity to mass ratio of
 \begin{equation} \left( \frac{L}{M} \right)_{\rm dyn} = \frac{v^{2}}{2\tau} =
 \frac{v^3\sigma}{2\lambda^3} \sim 8\e{-5}\, \frac{L_{\odot}}{M_{\odot}}, \end{equation}
 which is much smaller than the mass to luminosity relation implied by the model fit to
 the CO intensities ($\sim0.1$).  Lower densities and smaller clumps sizes mean more
 numerous clumps and a smaller mean separation.  This can increase the luminosity by a
 factor of only 10--20, but it is still several orders of magnitude too small.  Thus it is
 unlikely that a substantial fraction of the CO luminosity in the central $\sim$100~pc may
 be due to shocks from cloud-cloud collisions.  Again, while higher-velocity shocks could
 provide much more energy, and the $\sim$40~\kms\ non-circular velocities are suggestive
 that such shocks could exist, the H$_{2}$ ro-vibrational lines are inconsistent with any
 significant heating in fast shocks.  Thus while some amount of shock heating is required
 to produce the high SiO abundance, it is not likely to be a dominant factor in heating
 the bulk of the molecular gas.

\subsubsection{Cosmic Ray Heating}\label{sec:cosmic-ray-heating}
It is widely believed that low energy cosmic-ray protons dominate the heating of the quiescent,
UV-shielded molecular cloud cores in the disk of the Galaxy, keeping the gas temperature
above that of the CMB \citep[and references therein]{gl78}.  A much larger cosmic ray
density generated by massive star formation has also been invoked to explain the 50--150~K
measured molecular gas temperatures in the central regions of M82 (SAH).  Since the \hh\
cosmic ray dissociation cross sections are so small ($\sim 3\e{-26}\, \rm cm^{2}$), cosmic
rays easily penetrate molecular cloud cores.  The heating rate for cosmic rays can be
characterized by a ionization rate $\zeta_p$ per \hh\ molecule, and an average $\Delta Q$
deposited as thermal energy per ionization, that is: \begin{equation} \chi_{\rm H_2} =
\zeta_p\, \Delta Q. \end{equation} \citet{gl78} adopt values for $\Delta Q$ between 17 eV
and 20 eV.  For the ionization rate we solve Eq. 3 of SAH which scales the cosmic ray
density from that of the Galactic disk according to the supernova rate per unit area
$\psi$ and a typical wind velocity $v$ by which cosmic rays are removed from the disk of a
galaxy:
\begin{equation} 
\frac{\zeta_{p,\,\rm N253}}{\zeta_{p\,\rm Gal}} = \frac{n_{cr\,\rm N253}}{n_{cr,\,\rm Gal}}
\propto \frac{\psi_{\rm N253}}{\psi_{\rm Gal}}\,\frac{v_{\rm Gal disk}}{v_{w, \rm N253}}.
\end{equation}
The Galactic ionization rate $\zeta_{p,\rm\,Gal}$ is between $\rm 2\e{-17}\,s^{-1}$
\citep{gl78} and $\rm 7\e{-17}\,s^{-1}$ \citep{vb86}, (see also \citet{dot02}).
For the supernova rate in NGC~253 $\psi_{\rm N253}$, we take the lower limit derived for the central
360$\times$90 pc of 0.1 yr$^{-1}$ \citep{au88}, giving $3\rm yr^{-1}\,kpc^{-2}$.  To
compare with the Galaxy, we take a rate of 5.8 per century over the entire disk \citep[and
references therein]{vt91}, or $8\e{-5}\rm\, yr^{-1}\,kpc^{-2}$.  The diffusion velocity in
the Galactic disk is 10\kms\ (SAH), and the superwind velocity in NGC~253 is $\sim$500
\kms, \citep{str02}.  The ionization rate can then be scaled from that of the Galaxy:
\begin{equation}\rm \zeta_{p,\rm\, N253} =750\, \zeta_{p,\rm\,Gal} = 1.5-5.3\e{-14} s^{-1}. \end{equation} 
With a $\Delta Q$ of 20 eV per ionization, the power deposited per \hh\ molecule is then 
5--18~$\e{-25}\rm\,erg\,s^{-1}$.
This estimate represents a lower limit to the cosmic ray heating, yet it is very close to
the measured luminosity per \hh\ molecule in all of the CO lines,
$7\e{-25}\rm\,erg\,s^{-1}$ with our assumed CO abundance.

To check that the increased CR flux does not destroy \hh\ and other molecular
constituents, we compare our derived ionization rate with Eq. 10 of SAH.  This requires
that the ionization rate be less than the \hh\ formation rate:
\begin{equation}\rm \zeta_{\rm p\,N253} < 1.5\e{-18}\,n\,T^{1/2}, \end{equation} where
$\rm n$ is the
total hydrogen density.  For our parameters, this requires $\rm \zeta_{\rm p\,N253} <
7.4\e{-13}\, s^{-1}$, a factor of 50 from our lower limit of $\zeta_{\rm p\,N253}$.

It is conceivable the increased CR density would dissociate the CO in the cloud core.
\citet{far94} have constructed a chemical model of dark clouds to investigate the effects
of cosmic rays ionization on the ion-neutral chemistry.  For steady state models, with
cloud density $\sim 2\e{4}\, \rm cm^{-3}$ and visual extinction $A_{V} \sim 30$ they find
CO remains the dominant reservoir for carbon for the entire range of cosmic ray ionization
rates investigated, $\rm \zeta\, = 1.3\e{-19} \rightarrow 8\e{-14}\,s^{-1}$.  The neutral
carbon abundance grows sharply as $\zeta$ grows to $2\e{-14}$ approaching 25\% of the CO
abundance at the largest cosmic ray fluxes investigated.  These results are consistent
with the large $\rm C^{0}/CO$ abundance ratios observed in the nuclei of both NGC 253
\citep{har95}, and M82 \citep{sch93}.  Since cosmic ray heating provides the necessary
molecular gas luminosity and naturally accounts for the large observed neutral carbon
abundance, while other heating sources are insufficient by large factors, we favor favor
the cosmic rays as the dominant heating source for the bulk of the molecular gas in the
nucleus of NGC~253.  Our analysis is fairly general, and we note that similar arguments
could be made for other starburst nuclei where the rate of supernovae explosions
approaches 0.1 per year in a $\sim$100 pc region.  

\subsection{Cloud Size, Relation to Atomic Gas}\label{sec:cloud-size}

The total gas mass in a region compared with the derived density determines a volume
filling factor of clouds.  For our modeled dense component, the result is $\phi_V =
2.6\e{-3}$.  The physical size of the clouds, however, is not constrained by our
observations.  The existence of HCN in dense cores \citep{pag95} indicates that many
clouds must be large enough to shield their interiors from UV, approximately $\rm A_V >
10$ for the cloud radius.  This puts a lower limit on the cloud size according to
Eq~(\ref{eq:av}).  If we assume that the photodissociated gas, as traced through its [CII]
and [OI] line emission forms a surface layer on the molecular cloud interiors, then
we can estimate the cloud sizes.  As discussed in Section~\ref{sec:uv-energy-sources},
such a layer would have a thickness such that $\rm A_V \sim 3$, and the atomic to
molecular gas mass ratio requires that the atomic surface constitute $\sim 3\e{-2}$ of the
total cloud mass.  For spherical clouds, this implies a cloud radius corresponding to $\rm
A_V \sim 300$.  Converting to physical sizes, these arguments then give $\rm r_{cloud}
\sim 2.0\,\rm pc$, at our modeled density and solar dust abundance.  
In a plane-parallel geometry, appropriate for clouds illuminated from a single direction,
the geometric argument would suggest a cloud thickness corresponding to $\rm A_V \sim
100$, or 0.7~pc.  Adopting the 2~pc spherical clouds for argument, each has a mass of
$1.1\e{5}$ \ms, and there about 330 such clouds in the central 180~pc region.  For
reference, we note that increasing the CO abundance does not change the arguments above
regarding the individual cloud geometry, but would correspond to a smaller molecular gas
mass, lower volume filling factor, and fewer number of clouds.  Table~\ref{tab:props}
summarizes the parameters for our adopted model.

\section{Conclusions}

We have detected bright CO \jseven\ emission from the nucleus of NGC~253.  We combine our
observations with lower-J measurements of both $\rm ^{12}CO$ and $\rm ^{13}CO$ and an LVG
model to estimate the CO column density and physical conditions of the molecular gas in
the starburst nucleus.  The column density of CO we infer based on the transitions above
\jone\ is slightly larger than but similar to the estimates made by HHR99 for the central
23\as.  However, we find that the excitation conditions of this gas must be very high to
reproduce the observed \jsix\ and \jseven\ intensities.  A single component with $\rm T
\sim 120\, K$, and $\rm n\sim 4.5\e{4}\, cm^{-3}$ provides the best fit to all the
observed transitions above \jone\ for both $^{12}$CO and $^{13}$CO, and is consistent with
the ISO molecular \hh\ observations.  We adopt this as the favored model.  While a smaller
additional mass of cool gas is required to account for the $^{13}$CO \jone\ intensity, and
more cool gas may be present, the CO column density of warm molecular gas cannot be much
less than our derived values.

The large mass of warm gas is very difficult to explain with a photodissociation
region scenario.  The molecular gas mass is a factor of 10--30 greater than the atomic gas
mass, while the two should be comparable if heated with stellar UV as in the Galactic PDR
templates.  Dynamical heating of molecular gas from shocks is also unlikely based on
energetic considerations, and given the lack of evidence for significant heating in fast
shocks.  On the other hand, the very large cosmic ray flux in the starburst nucleus of
NGC~253 readily provides the necessary energy to heat the gas, and a means to deliver
energy throughout the volume of large, UV-shielded clouds.  It is likely the
dominant heating term in the molecular ISM energy budget.  While some of the warm
molecular gas that we observe is likely heated by the far-UV radiation and shocks,
our results strongly support the conjecture by \citep{suc93} that the bulk of molecular
gas in starburst nuclei is held at high excitation by the enhanced cosmic ray flux.  

\section {Acknowledgments}
We are very grateful to the staff of JCMT for their support and patience during the first
SPIFI observing run.  We especially thank the director of the JCMT, Ian Robson for his
support, and Wayne Holland for his tireless efforts in SPIFI's setup and first
observations.  We are also indebted to Cornell engineer Chuck Henderson for solutions to
difficult problems at the telescope on the instrument's first observing run.  We thank an
anonymous referee for several helpful suggestions with this manuscript.  This
work was supported by NASA grants NAGW-4503 and NAGW-3925, and NSF grants OPP-8920223, and
OPP-0085812.

%
%
 
\clearpage 

\end{document}